\documentclass[12pt]{article}
\usepackage{amsfonts}

\title{Spontaneous Magnetization in the Finite XXZ Spin Chain with Boundaries}
\author{Yasuhiro Fujii\footnotemark[2]\ \ and Miki Wadati\footnotemark[3]}
\date{Department of Physics, Graduate School of Science, \\
  University of Tokyo,\\  Hongo 7-3-1, Bunkyo-ku, Tokyo 113-0033, Japan}

\begin{document}

\maketitle
\baselineskip 5.3mm

\footnotetext[2]{e-mail: \texttt{fujii@monet.phys.s.u-tokyo.ac.jp}}
\footnotetext[3]{e-mail: \texttt{wadati@monet.phys.s.u-tokyo.ac.jp}}

\begin{abstract}
  The finite XXZ spin chain with boundaries is studied.
  We derive the transfer matrix from
  the $q$-difference equation discovered by Cherednik
  and construct its eigenstates by the vertex operator approach.
  We point out that the eigenstates with no magnetic fields
  have a symmetry called the turning symmetry.
  Making use of this symmetry
  we calculate the spontaneous magnetization
  in the thermodynamic limit,
  which is roughly twice as large as that
  in the half-infinite XXZ spin chain.  
\end{abstract}

\section{Introduction}
The XXZ spin chain is the fundamental model
in the study of the integrable systems.
It satisfies the Yang-Baxter equation,
which enables us to clarify exactly
the structure of the energy levels.
Further, the Yang-Baxter equation yields an realization of quantum group
and nowadays its representation theory promotes
the analysis of various integrable models
\cite{CBMS,JMMN,DFJMN}.

Encouraged by the success of the conformal field theory \cite{BPZ,KZ},
much attention has been focused
also to the massive integrable field theories.
A remarkable result is the discovery of the $q$-difference equation
similar to the Knizhnik-Zamolodchikov equation \cite{FR},
which is satisfied by the correlation functions
in the integrable models.
This equation is called the quantum Knizhnik-Zamolodchikov (q-KZ) equation
and has been studied extensively in both mathematics and physics.
Solutions of the q-KZ equation are formulated with various methods,
for examples, the Jackson integral \cite{Ma,V},
the algebraic Bethe ansatz \cite{R}
and the vertex operators (VOs) in the quantum affine algebra
\cite{CBMS,DFJMN,KQS}.
For instance,
the correlation functions in the infinite XXZ spin chain
are constructed by the VOs
and satisfy the q-KZ equation.
It is also known that the Smirnov's form factor is one of solutions of
the q-KZ equation \cite{FR,Sm}.

The above discussions can be extended to the half-infinite chain
with a special boundary:\
there exist a boundary condition
that preserves the integrability \cite{Sk}.
Such a boundary condition is described
by the boundary K-matrix satisfying the reflection equation
instead of the Yang-Baxter equation.
Using this K-matrix
and noting that the VOs physically correspond to the half-infinite chain
\cite{CBMS},
we can formulate correlation functions
in the half-infinite XXZ spin chain
in terms of the VOs \cite{JKKKM1}.
Moreover,
they satisfy the $q$-difference equation
similar to the q-KZ equation \cite{C}.
In the paper this equation is called the boundary q-KZ equation.
By manipulation of the boundary q-KZ equation,
in the half-infinite XXZ spin chain,
the two-point function is exactly calculated \cite{JKKKM2}
and by the explicit representations of the VOs
the correlation functions are found in the integral expressions
\cite{JKKKM1}.

Our purpose in this paper is to calculate the correlation functions
of the finite XXZ spin chain with boundaries
by the vertex operator approach.
It is obviously impossible
to analyze the finite chain in the same manner as above,
because the VO implies the half-infinite chain.
However it has been believed that, in the thermodynamic limit,
the bounded chain is substituted for the half-infinite chain,
because the vertex operator approach gives the same result as
that obtained by analysis of the Bethe equation for a finite chain
\cite{CBMS,DFJMN,BVV}.
Can it be true?
We bring forward a counterargument to this expectation
by noting special bounded spin chains
such that they are invariant under overturning
(called the turning symmetry).
The half-infinite chain never satisfies such geometric symmetry.
The turning symmetry may strongly influence
physical behaviors of the model.

To treat the finite XXZ spin chain
we present a new idea.
We pay attention to the resemblance between
the boost operator in the boundary q-KZ equation
and the transfer matrix of the bounded model.
In section \ref{bqkz}
we introduce the boundary q-KZ equation and
get the transfer matrix from this equation.
Then the eigenstates of the model is made of the VOs.
In section \ref{energy}
we introduce the idea of the turning symmetry
and compute the vacuum energy
making use of this symmetry.
We also construct the excited states using the VOs of another type
and compute these energy levels.
It is found that
the structure of energies is similar to the quasi-particle
structure pointed out by Faddeev and Takhtajan in the XXX model \cite{FT}.
In section \ref{thermo} the thermodynamic limit is considered.
The vacuum is then expressed by the infinite product of VOs.
We show that
such infinite product is interpreted as a huge group of R-matrices
and is simplified by the turning symmetry.
Its asymptotic form is obtained
by applying the Baxter's formula on the corner transfer matrix \cite{B}.
In section \ref{spon} the spontaneous magnetization at a boundary
is calculated by the free field representations of the VOs.
We find that
it is larger than that in the half-infinite XXZ chain.

\setcounter{equation}{0}
\section{Boundary q-KZ Equation and Transfer Matrix}
\label{bqkz}
In this section we introduce the $q$-difference equation
called the boundary q-KZ equation
and solve it by the vertex operator approach.
From the boundary q-KZ equation
the transfer matrix in the finite XXZ spin chain
is squeezed out.
Then the solution of the boundary q-KZ equation
corresponds to the eigenstate of the model,
which is expressed in terms of the VOs.

\subsection{Boundary q-KZ equation}
As an extension of the q-KZ equation
Cherednik proposed the following difference equation \cite{C},
\begin{equation}
  \label{BqKZ}
  F(\zeta_1,\ldots,q^{-2}\zeta_j,\ldots,\zeta_N)
  =
  \bar{T}_j(\zeta_1,\ldots,\zeta_N) F(\zeta_1,\ldots,\zeta_N),
\end{equation}
where $\bar{T}_j(\zeta_1,\ldots,\zeta_N)$ is the boost operator given by
\begin{eqnarray}
  \label{Tj}
  \bar{T}_j(\zeta_1,\ldots,\zeta_N)
  &=&
  R_{j,j-1}(\zeta_j/q^2\zeta_{j-1})\cdots R_{j,1}(\zeta_j/q^2\zeta_1)
  \bar{K}_j(\zeta_j)
  \nonumber \\
  && \times
  R_{1,j}(\zeta_1 \zeta_j)\cdots R_{j-1,j}(\zeta_{j-1}\zeta_j)
  R_{j+1,j}(\zeta_{j+1}\zeta_j)\cdots R_{N,j}(\zeta_N \zeta_j)
  \nonumber \\
  && \times
  K_j(\zeta_j)R_{j,N}(\zeta_j/\zeta_N)\cdots R_{j,j+1}(\zeta_j/\zeta_{j+1}).
\end{eqnarray}
The R-matrix $R_{i,j}(\zeta)$ is the intertwiner of $U_q(\widehat{sl}_2)$
for a map $\mbox{End}_\mathbb{C}(V_i\otimes V_j)$.
$K_j(\zeta)$ and $\bar{K}_j(\zeta)$ are the boundary K-matrices
defined by a map $\mbox{End}_\mathbb{C}(V_j)$ \cite{Sk}.
They are parameterized by $r$ and $\bar{r}$ such that
$K_j(\zeta) = K_j(\zeta;r)$ and
$\bar{K}_j(\zeta) = K_j(\zeta;\bar{r})$, respectively.
Let $v_\epsilon$ ($\epsilon=\pm$) denote the natural basis of
$V=\mathbb{C}^2$.
The matrix elements of $R(\zeta)$ and $K(\zeta;r)$
are defined by the relations,
\begin{eqnarray}
  R(\zeta).(v_{\epsilon_1}\otimes v_{\epsilon_2})
  &=&
  \sum_{\epsilon'_1, \epsilon'_2=\pm}
  R_{\epsilon_1 \epsilon_2}^{\epsilon'_1 \epsilon'_2}(\zeta)
  (v_{\epsilon'_1}\otimes v_{\epsilon'_2}),
  \\
  K(\zeta;r).v_\epsilon
  &=&
  \sum_{\epsilon'=\pm} K_\epsilon^{\epsilon'}(\zeta;r) v_{\epsilon'}.
\end{eqnarray}
The solution $F(\zeta_1,\ldots,\zeta_N)$ forms an $N$-fold tensor space
$V_1\otimes\cdots\otimes V_N$.
It was shown \cite{JKKKM2} that the solution
of equation (\ref{BqKZ}) corresponds to
the correlation function in the half-infinite XXZ spin chain
with a boundary.
In the paper we call the equation (\ref{BqKZ}) the boundary q-KZ equation.

The half-infinite XXZ model is formulated in terms of the VOs
of the quantum affine algebra $U_q(\widehat{sl}_2)$ \cite{JKKKM1,JKKKM2}.
To solve the boundary q-KZ equation
we need the VO
$\Phi(\zeta) = \sum_{\epsilon=\pm}\Phi_\epsilon(\zeta).v_\epsilon$,
the dual VO $\Phi^*(\zeta)$
related with $\Phi_\epsilon^*(\zeta)=\Phi_{-\epsilon}(-q^{-1}\zeta)$
and some states $|W\rangle$, $\langle W^*|$
defined by the following relations,
\begin{eqnarray}
  \label{W}
  \sum_{\epsilon'} K_\epsilon^{\epsilon'}(\zeta)
  \Phi_{\epsilon'}(\zeta)|W\rangle
  &=&
  \Phi_\epsilon(\zeta^{-1})|W\rangle,
  \\
  \label{W*}
  \sum_{\epsilon'} \langle W^*| \Phi_{\epsilon'}^*(\zeta^{-1})
  \bar{K}_{\epsilon'}^\epsilon(\zeta)
  &=&
  \langle W^*| \Phi_\epsilon^*(\zeta).
\end{eqnarray}
Then the solution of the boundary q-KZ equation is written as
\begin{eqnarray}
  F(\zeta_1,\ldots,\zeta_N)
  &=&
  \langle W^*| \Phi(\zeta_1)\otimes\cdots\otimes\Phi(\zeta_N) |W\rangle
  \nonumber \\
  &=&
  \sum_{\epsilon_1,\ldots,\epsilon_N}
  \langle W^*| \Phi_{\epsilon_1}(\zeta_1)\cdots\Phi_{\epsilon_N}(\zeta_N)
  |W\rangle (v_{\epsilon_1}\otimes\cdots\otimes v_{\epsilon_N}).
\end{eqnarray}
The proof is readily done.
Noting that the commutation relation of the VOs,
\begin{equation}
  \label{com_VO}
  \sum_{\epsilon'_1,\epsilon'_2}
  R_{\epsilon_1 \epsilon_2}^{\epsilon'_1 \epsilon'_2}(\zeta_1/\zeta_2)
  \Phi_{\epsilon'_1}(\zeta_1) \Phi_{\epsilon'_2}(\zeta_2)
  =
  \Phi_{\epsilon_2}(\zeta_2) \Phi_{\epsilon_1}(\zeta_1),
\end{equation}
and the definition of the states $|W\rangle$, $\langle W^*|$,
one can check that $F(\zeta_1\ldots,\zeta_N)$ satisfies
the following relations,
\begin{eqnarray}
  R_{j,j+1}(\zeta_j/\zeta_{j+1}) F(\zeta_1,\ldots,\zeta_N)
  &=&
  F(\zeta_1,\ldots,\zeta_{j+1},\zeta_j,\ldots,\zeta_N),
  \\
  K_N(\zeta_N) F(\zeta_1,\ldots,\zeta_N)
  &=&
  F(\zeta_1,\ldots,\zeta_N^{-1}),
  \\
  \bar{K}_1(\zeta_1) F(\zeta_1^{-1},\ldots,\zeta_N)
  &=&
  F(q^{-2}\zeta_1,\ldots,\zeta_N),
\end{eqnarray}
where the order of the spectral parameters corresponds to
that of $N$-fold vector spaces.
These relations show that
\begin{eqnarray}
  \lefteqn{\bar{T}_j(\zeta_1,\ldots,\zeta_N) F(\zeta_1,\ldots,\zeta_N)}
  \nonumber \\
  &=&
  R_{j,j-1}(\zeta_j/q^2\zeta_{j-1})\cdots R_{j,1}(\zeta_j/q^2\zeta_1)
  \bar{K}_j(\zeta_j)
  \nonumber \\
  && \times
  R_{1,j}(\zeta_1 \zeta_j)\cdots R_{j-1,j}(\zeta_{j-1}\zeta_j)
  R_{j+1,j}(\zeta_{j+1}\zeta_j)\cdots R_{N,j}(\zeta_N \zeta_j)
  \nonumber \\
  && \times
  K_j(\zeta_j)R_{j,N}(\zeta_j/\zeta_N)\cdots R_{j,j+1}(\zeta_j/\zeta_{j+1})
  \nonumber \\
  && \times
  F(\zeta_1,\ldots,\zeta_N)
  \nonumber \\
  &=&
  R_{j,j-1}(\zeta_j/q^2\zeta_{j-1})\cdots R_{j,1}(\zeta_j/q^2\zeta_1)
  \bar{K}_j(\zeta_j)
  \nonumber \\
  && \times
  R_{1,j}(\zeta_1 \zeta_j)\cdots R_{j-1,j}(\zeta_{j-1}\zeta_j)
  R_{j+1,j}(\zeta_{j+1}\zeta_j)\cdots R_{N,j}(\zeta_N \zeta_j)
  \nonumber \\
  && \times
  F(\zeta_1,\ldots,\zeta_{j-1},\zeta_{j+1},\ldots,\zeta_N,\zeta_j^{-1})
  \nonumber \\
  &=&
  R_{j,j-1}(\zeta_j/q^2\zeta_{j-1})\cdots R_{j,1}(\zeta_j/q^2\zeta_1)
  \bar{K}_j(\zeta_j)
  \nonumber \\
  && \times
  F(\zeta_j^{-1},\zeta_1,\ldots,\zeta_{j-1},\zeta_{j+1},\ldots,\zeta_N)
  \nonumber \\
  &=&
  F(\zeta_1,\ldots,\zeta_{j-1},q^{-2}\zeta_j,\zeta_{j+1},\ldots,\zeta_N),
\end{eqnarray}
which completes the proof.
The boundary q-KZ equation is useful
in calculating the two-point correlation functions
not only in the XXZ model but in the XYZ model \cite{JKKKM2}.

\subsection{Transfer Matrix in the Finite XXZ Spin Chain}
One may notice that the boost operator (\ref{Tj})
is similar to the transfer matrix
of the XXZ model.
Now we consider the case such that the parameter $q^{-2}$ does not appear,
\begin{eqnarray}
  \label{T}
  T_j(\zeta_1,\ldots,\zeta_N)
  &=&
  R_{j,j-1}(\zeta_j/\zeta_{j-1})\cdots R_{j,1}(\zeta_j/\zeta_1)
  \bar{K}_j(\zeta_j)
  \nonumber \\
  && \times
  R_{1,j}(\zeta_1 \zeta_j)\cdots R_{j-1,j}(\zeta_{j-1}\zeta_j)
  R_{j+1,j}(\zeta_{j+1}\zeta_j)\cdots R_{N,j}(\zeta_N \zeta_j)
  \nonumber \\
  && \times
  K_j(\zeta_j)R_{j,N}(\zeta_j/\zeta_N)\cdots R_{j,j+1}(\zeta_j/\zeta_{j+1}).
\end{eqnarray}
This operator is obtained when the defining relation (\ref{W*}) is
modified (see (\ref{V})).
It is expected that the solution of the boundary q-KZ equation
for this operator gives the eigenstate of the XXZ model.
First we show that the operator (\ref{T}) can be regarded
as the transfer matrix for the finite XXZ spin chain.
We write the matrix elements of the R-matrix as follows,
\begin{eqnarray}
  &&
  R_{++}^{++}(\zeta) = R_{--}^{--}(\zeta) = \gamma(\zeta), \\
  &&
  R_{+-}^{+-}(\zeta) = R_{-+}^{-+}(\zeta)
  = \frac{(1-\zeta^2)q}{1-q^2\zeta^2}\gamma(\zeta), \\
  &&
  R_{+-}^{-+}(\zeta) = R_{-+}^{+-}(\zeta)
  = \frac{(1-q^2)\zeta}{1-q^2\zeta^2}\gamma(\zeta),
\end{eqnarray}
with the normalization factor $\gamma(\zeta)$
\begin{equation}
  \gamma(\zeta) =
  \frac{1}{\zeta}
  \frac{(q^2\zeta;q^4)_\infty (q^4\zeta^{-2};q^4)_\infty}
  {(q^2\zeta^{-2};q^4)_\infty (q^4\zeta^2;q^4)_\infty},
  \qquad
  (z;p)_\infty =
  \prod_{n=0}^\infty(1-zp^n).
\end{equation}
Other elements are $0$.
The normalization factor $\gamma(\zeta)$ is derived
from the commutation relation of the VOs (\ref{com_VO}).
The R-matrix satisfies the Yang-Baxter equation,
the unitarity relation and the crossing symmetry,
\begin{eqnarray}
  \label{YBE}
  &&
  R_{12}(\zeta_1/\zeta_2)R_{13}(\zeta_1/\zeta_3)R_{23}(\zeta_2/\zeta_3)
  =
  R_{23}(\zeta_2/\zeta_3)R_{13}(\zeta_1/\zeta_3)R_{12}(\zeta_1/\zeta_2),
  \\ 
  && 
  R_{12}(\zeta_1/\zeta_2)R_{21}(\zeta_2/\zeta_1) = 1,
  \\
  \label{cross}
  && 
  R_{\epsilon_1 \epsilon_2}^{\epsilon'_1 \epsilon'_2}(\zeta_1/\zeta_2)
  =
  R_{\epsilon_2 -\epsilon'_1}^{\epsilon'_2 -\epsilon_1}
  (\zeta_2/q\zeta_1).
\end{eqnarray}
The K-matrix is given by the following diagonal matrix,
\begin{equation}
  K_+^+(\zeta;r) = \frac{1-r\zeta^2}{\zeta^2-r}\kappa(\zeta;r),
  \qquad
  K_-^-(\zeta;r) = \kappa(\zeta;r),
\end{equation}
\begin{equation}
  \kappa(\zeta;r) =
  \frac{(q^4 r\zeta^2;q^4)_\infty (q^2 r\zeta^{-2};q^4)_\infty
    (q^6 \zeta^4;q^8)_\infty (q^8 \zeta^{-4};q^8)_\infty}
  {(q^4 r\zeta^{-2};q^4)_\infty (q^2 r\zeta^2;q^4)_\infty
    (q^6 \zeta^{-4};q^8)_\infty (q^8 \zeta^4;q^8)_\infty}.
\end{equation}
The K-matrix satisfies the reflection equation,
the boundary unitarity relation and the boundary crossing symmetry,
\begin{eqnarray}
  &&
  K_2(\zeta_2)R_{21}(\zeta_1\zeta_2)K_1(\zeta_1)R_{12}(\zeta_1/\zeta_2)
  =
  R_{21}(\zeta_1/\zeta_2)K_1(\zeta_1)R_{12}(\zeta_1\zeta_2)K_2(\zeta_2),
  \\
  &&
  K(\zeta)K(\zeta^{-1}) = 1,
  \\
  \label{b_cross}
  &&
  K_{\epsilon_1}^{\epsilon_2}(\zeta)
  =
  \sum_{\epsilon'_1, \epsilon'_2}
  R_{-\epsilon'_2 \epsilon_1}^{\epsilon_2 -\epsilon'_1}(\zeta^2)
  K_{\epsilon'_2}^{\epsilon'_1}(-q^{-1}\zeta^{-1}).  
\end{eqnarray}
The normalization factor $\kappa(\zeta)$ is determined by
the boundary crossing symmetry (\ref{b_cross}).
The above six relations for the R-matrix and the K-matrix play
important roles in solving the model exactly.
From these explicit representations
those derivatives are
\begin{eqnarray}
  \left.\frac{\partial}{\partial\zeta}
    R_{j,j+1}(\zeta)P_{j,j+1}\right|_{\zeta=1}
  &=&
  -\frac{q}{1-q^2}
  (\sigma_j^x \sigma_{j+1}^x + \sigma_j^y \sigma_{j+1}^y
  + \Delta \sigma_j^z \sigma_{j+1}^z) + \mbox{const.},
  \\
  \left.\frac{\partial}{\partial\zeta}K_{j}(\zeta;r)\right|_{\zeta=1}
  &=&
  -\frac{1+r}{1-r}\sigma_j^z + \mbox{const.},
\end{eqnarray}
where $\sigma^x$, $\sigma^y$, $\sigma^z$ are the Pauli matrices
and the anisotropic parameter $\Delta=(q+q^{-1})/2$.
$P_{i,j}$ is the permutation operator:
$V_i\otimes V_j\rightarrow V_j\otimes V_i$.
In the paper we work in the antiferromagnetic region $\Delta<-1$
or $-1<q<0$.
Taking notice of the relation $R_{j,j+1}(1)=P_{j,j+1}$ and
$R_{j,j+1}(\zeta) P_{j,i} = P_{j,i} R_{i,j+1}(\zeta)$,
we get the derivative of $T(\zeta_1,\ldots,\zeta_N)$ as follows,
\begin{eqnarray}
  \lefteqn{\left.
      \frac{\partial}{\partial\zeta_j} T_j(\zeta_1,\ldots,\zeta_N)
    \right|_{\zeta_1=\cdots=\zeta_N=1}}
  \nonumber \\
  &=&
  \sum_{i=1}^{j-1} \left.
    \frac{\partial}{\partial\zeta_j} R_{i+1,i}(\zeta_j)P_{i+1,i}
  \right|_{\zeta_j=1}
  + \left.
    \frac{\partial}{\partial\zeta_j} \bar{K}_1(\zeta_j)\right|_{\zeta_j=1}
  + \sum_{i=1}^{j-1} \left.
    \frac{\partial}{\partial\zeta_j} R_{i,i+1}(\zeta_j)P_{i,i+1}
  \right|_{\zeta_j=1}
  \nonumber \\
  &&
  + \sum_{i=j}^{N-1} \left.
    \frac{\partial}{\partial\zeta_j} R_{i+1,i}(\zeta_j)P_{i+1,i}
  \right|_{\zeta_j=1}
  + \left.
    \frac{\partial}{\partial\zeta_j} K_N(\zeta_j)\right|_{\zeta_j=1}
  + \sum_{i=j}^{N-1} \left.
    \frac{\partial}{\partial\zeta_j} R_{i,i+1}(\zeta_j)P_{i,i+1}
  \right|_{\zeta_j=1}
  \nonumber \\
  &=&
  \frac{4q}{1-q^2} H_{\mbox{\scriptsize XXZ}} + \mbox{const.}.
\end{eqnarray}
Here $H_{\mbox{\scriptsize XXZ}}$ is the Hamiltonian
of the XXZ model with boundary magnetic fields,
\begin{equation}
  H_{\mbox{\scriptsize XXZ}} =
  -\frac{1}{2} \sum_{i=1}^{N-1}
  (\sigma_i^x \sigma_{i+1}^x + \sigma_i^y \sigma_{i+1}^y
  + \Delta \sigma_i^z \sigma_{i+1}^z)
  + h_1\sigma_1^z + h_N\sigma_N^z,
\end{equation}
and the boundary magnet fields $h_1$, $h_N$ are related
with parameters $\bar{r}$, $r$, 
\begin{equation}
  \label{h}
  h_1 = -\frac{1-q^2}{4q}\frac{1+\bar{r}}{1-\bar{r}},
  \qquad
  h_N = -\frac{1-q^2}{4q}\frac{1+r}{1-r}.
\end{equation}
We thus have shown that the operator (\ref{T})
generates the transfer matrix for the finite XXZ spin chain.

To get this transfer matrix
we modify the relation (\ref{W*}) into
\begin{equation}
  \label{V}
  \sum_{\epsilon'} \langle V|
  \Phi_{\epsilon'}(\zeta^{-1}) \bar{K}_{\epsilon'}^\epsilon(\zeta)
  =
  \langle V| \Phi_\epsilon(\zeta).
\end{equation}
We call the states $|W\rangle$ and $\langle V|$
the boundary states.
Then the boundary q-KZ equation is written as
\begin{equation}
  T_j(\zeta_1,\ldots,\zeta_N) |\zeta_1,\ldots,\zeta_N\rangle
  =
  |\zeta_1,\ldots,\zeta_N\rangle,
\end{equation}
with the state $|\zeta_1,\ldots,\zeta_N\rangle$ being defined by
\begin{eqnarray}
  |\zeta_1,\ldots,\zeta_N\rangle
  &=&
  \langle V|
  \Phi(\zeta_1)\otimes\cdots\otimes\Phi(\zeta_N)
  |W\rangle
  \nonumber \\
  &=&
  \sum_{\epsilon_1,\ldots,\epsilon_N}
  \langle V|
  \Phi_{\epsilon_1}(\zeta_1)\cdots\Phi_{\epsilon_N}(\zeta_N)
  |W\rangle
  (v_{\epsilon_1}\otimes\cdots\otimes v_{\epsilon_N}).
\end{eqnarray}
Note that the state $|\zeta_1,\ldots,\zeta_N\rangle$ is
in an $N$-fold tensor space
$V_1\otimes\cdots\otimes V_N$.
We regard this state as the eigenstate of the transfer matrix
of the XXZ model.

We point out a similarity between such construction
by the boundary q-KZ and the matrix product ansatz
in the one-dimensional reaction-diffusion processes with boundaries
\cite{DEHP}.
In the matrix product ansatz the eigenstate is
expressed by a product of the Zamolodchikov-Faddeev algebra,
which is determined by the ``Hamiltonian'' describing
the time-development of the model \cite{SW}.
In the viewpoint of the matrix product ansatz,
the VO is a kind of the Zamolodchikov-Faddeev algebra,
since it obeys the same commutation relation (\ref{com_VO}).
However the original matrix product ansatz is not applicable
to the XXZ model,
because the structure of the energy levels may be different.
Details are discussed in the next section.
Our method to construct the eigenstates is restricted to
the integrable models whose correlation functions satisfy
$q$-difference equations like the q-KZ equation.

\setcounter{equation}{0}
\section{Energy Levels and Turning Symmetry} 
\label{energy}
Strictly speaking,
there exist no boundary states satisfying the defining relations
(\ref{W}) and (\ref{V}),
which include some extra factors
$\Lambda_W(\zeta;r)$ and $\Lambda_V(\zeta;r)$.
These factors determine the vacuum energy of the model.
In this section we slightly modify the boundary states
from a mathematical standpoint
and also construct the excited states.

The quantum affine algebra $U_q(\widehat{sl}_2)$
has two types of the VOs.
The VO used in the previous section $\Phi(\zeta)$
is the type I, which is defined by
the intertwiner for a map
$\Phi^{(1-i,i)}(\zeta):
V(\Lambda_i) \rightarrow V(\Lambda_{1-i})\otimes V$ ($i=0,1$).
Here $V(\Lambda_i)$ is the highest weight representations
of $U_q(\widehat{sl}_2)$ with the level $1$.
This mathematical definition leads to
\begin{equation}
  |W\rangle^{(i)} \in V(\Lambda_i),
  \qquad
  {}^{(i)}\langle V| \in V^*(\Lambda_i),
\end{equation}
with the dual highest weight representations $V^*(\Lambda_i)$.
Using scalar factors $\Lambda_W^{(i)}(\zeta;r)$ and $\Lambda_V^{(i)}(\zeta;r)$
we generalize the relations (\ref{W}) and (\ref{V}) into
\begin{eqnarray}
  \label{LW}
  \sum_{\epsilon'} K_\epsilon^{\epsilon'}(\zeta)
  \Phi_{\epsilon'}^{(1-i,i)}(\zeta) |W\rangle^{(i)}
  &=&
  \Lambda_W^{(i)}(\zeta;r) \Phi_\epsilon^{(1-i,i)}(\zeta^{-1})
  |W\rangle^{(i)},
  \\
  \label{LV}
  \sum_{\epsilon'} {}^{(i)}\langle V|
  \Phi_{\epsilon'}^{(i,1-i)}(\zeta^{-1}) K_{\epsilon'}^\epsilon(\zeta)
  &=&
  \Lambda_V^{(i)}(\zeta;\bar{r}) {}^{(i)}\langle V|
  \Phi_\epsilon^{(i,1-i)}(\zeta).
\end{eqnarray}
These scalar factors are given by
(see Section \ref{bound})
\begin{equation}
  \Lambda_W^{(0)}(\zeta;r) = \Lambda_V^{(1)}(\zeta;r) = 1,
  \qquad
  \Lambda_W^{(1)}(\zeta;r) = \Lambda_V^{(0)}(\zeta;r) = \Lambda(\zeta;r),
\end{equation}
where
\begin{equation}
  \Lambda(\zeta;r) =
  \frac{1}{\zeta^2}
  \frac{\Theta_{q^4}(r\zeta^2)\Theta_{q^4}(q^2r\zeta^{-2})}
  {\Theta_{q^4}(r\zeta^{-2})\Theta_{q^4}(q^2r\zeta^2)}.
\end{equation}
Using the boundary states $|W\rangle^{(i)}$ and ${}^{(i)}\langle V|$,
we redefine the eigenstates of the transfer matrix by
\begin{eqnarray}
  \hspace*{-1em}
  |\zeta_1,\ldots,\zeta_N\rangle^{(i)}
  &=&
  {}^{(i)}\langle V|
  \Phi^{(i,1-i)}(\zeta_1)\otimes\cdots\otimes\Phi^{(1-i,i)}(\zeta_N)
  |W\rangle^{(i)}
  \nonumber \\
  &=&
  \sum_{\epsilon_1,\ldots,\epsilon_N}
  {}^{(i)}\langle V|
  \Phi_{\epsilon_1}^{(i,1-i)}(\zeta_1)\cdots
  \Phi_{\epsilon_N}^{(1-i,i)}(\zeta_N)
  |W\rangle^{(i)}
  (v_{\epsilon_1}\otimes\cdots\otimes v_{\epsilon_N}).
\end{eqnarray}
The labels $(1-i,i)$ decorating the VOs are omitted
if there is no danger of confusion.

Suppose that the number of sites $N$ is even.
We turn the eigenstates upside down,
which means that the spins and the their orders are reversed.
With this operation we have
\begin{eqnarray}
  \lefteqn{\hspace*{-4em}
    \sum_{\epsilon_1,\ldots,\epsilon_N}
    {}^{(i)}\langle V(r^{-1})|
    \Phi_{-\epsilon_N}(\zeta_N)\cdots\Phi_{-\epsilon_1}(\zeta_1)
    |W(\bar{r}^{-1})\rangle^{(i)}
    (v_{-\epsilon_N}\otimes\cdots\otimes v_{-\epsilon_1})}
  \nonumber \\
  & \simeq &
  {}^{(i)}\langle V(r^{-1})|
  \Phi(\zeta_1)\otimes\cdots\otimes\Phi(\zeta_N)
  |W(\bar{r}^{-1})\rangle^{(i)}.
\end{eqnarray}
Here we have denoted the parameter $r$ dependence explicitly.
Pay attention to the defining relations of
the boundary magnetic fields (\ref{h}):\
the parameter $r$ is inverted
when the direction of the magnetic field is reversed.
If the eigenstates are equivalent to the states
turned upside down,
the relation $\bar{r} = r^{-1}$ is derived.
We call such geometric symmetry the \textit{turning symmetry}.
Since the spontaneous magnetizations are
generated by the states satisfying the turning symmetry,
we hereafter consider only the states whose number of sites is even
and which satisfy the turning symmetry.
Then the eigenvalue $\Lambda^{(i)}(\zeta_j;r)$
of the transfer matrix $T_j(\zeta_1,\ldots,\zeta_N)$
with the eigenstate $|\zeta_1,\ldots,\zeta_N\rangle^{(i)}$ are
\begin{equation}
  \Lambda^{(0)}(\zeta;r) = \Lambda(\zeta;r^{-1}) = \Lambda(\zeta;r)^{-1},
  \qquad
  \Lambda^{(1)}(\zeta;r) = \Lambda(\zeta;r).
\end{equation}
Noting that $T_j(1,\ldots,1)=\Lambda^{(i)}(1;r)=1$,
differentiating $T_j(\zeta_1,\ldots,\zeta_N)$ and $\Lambda^{(i)}(\zeta;r)$
at $\zeta_1=\cdots=\zeta_N=1$, 
we have the following energy spectrum,
\begin{equation}
  H_{\mbox{\scriptsize XXZ}}|1,\ldots,1 \rangle^{(i)} =
  (-1)^{1-i}\epsilon(r) |1,\ldots,1 \rangle^{(i)},
\end{equation}
\begin{equation}
  \epsilon(r) =
  \left\{
    \begin{array}{cc}
      \displaystyle{\frac{2Kk'}{\pi}\sinh\frac{\pi K'}{K}}
      \mbox{sn}(2K'\alpha) & (-1<r<0) \\
      \displaystyle{\frac{2K}{\pi}\sinh\frac{\pi K'}{K}}
      \frac{1}{\mbox{sn}(2K'\alpha)}. & (0<r<1) 
    \end{array}
  \right.
\end{equation}
Here $k'$ is the complementary modulus of the elliptic function
and a common constant factor is omitted.
$r$ and $q$ are parameterized as
\begin{equation}
  r = \left\{
    \begin{array}{cc}
      -(q^2)^\alpha & (-1<r<0) \\
      (q^2)^\alpha & (0<r<1)
    \end{array}
  \right.,
  \qquad
  q = -e^{-\pi K'/K}.
\end{equation}
From this energy spectrum
the vacuum state is determined as follows,
\begin{equation}
  \label{vac}
  |\mbox{vac}\rangle =
  \left\{
    \begin{array}{cl}
      |\zeta_1,\ldots,\zeta_N \rangle^{(0)} & (|r|<1) \\
      |\zeta_1,\ldots,\zeta_N \rangle^{(1)} & (|r|>1).
    \end{array}
  \right.
\end{equation}
To assure that the state (\ref{vac}) really defines the vacuum,
we show that
any excited states can be constructed
by the type II VO $\Psi_\mu^*(\zeta)$ ($\mu=\pm$)
in the same way as the infinite XXZ spin chain
\cite{CBMS,DFJMN}.
We introduce the $M$-particles excited states by
\begin{eqnarray}
  \lefteqn{|\zeta_1,\ldots,\zeta_N ; \xi_1,\ldots,\xi_M
    \rangle^{(i)}_{\mu_1,\ldots,\mu_M}}
  \nonumber \\
  &=&
  {}^{(i)}\langle V|
  \Phi(\zeta_1)\otimes\cdots\otimes \Phi(\zeta_N)
  \Psi_{\mu_1}^*(\xi_1)\cdots \Psi_{\mu_M}^*(\xi_M)
  |W\rangle^{(i)},
\end{eqnarray}
and confirm that this state corresponds to the excited state.
Using the communication relation between
the type I and type II VOs,
\begin{equation}
  \Phi_\epsilon(\zeta) \Psi_\mu(\xi)
  =
  \tau(\zeta/\xi)\Psi_\mu(\xi) \Phi_\epsilon(\zeta),
\end{equation}
\begin{equation}
  \tau(\zeta) =
  \frac{1}{\zeta}\frac{\Theta_{q^4}(q\zeta^2)}{\Theta_{q^4}(q\zeta^{-2})},
  \qquad
  \Theta_p(z) =
  (z;p)_\infty (pz^{-1};p)_\infty (p;p)_\infty,
\end{equation}
we obtain the eigenvalues,
\begin{eqnarray}
  \lefteqn{T_j(\zeta_1,\ldots,\zeta_N)
    |\zeta_1,\ldots,\zeta_N ; \xi_1,\ldots,\xi_M \rangle^{(i)}}
  \nonumber \\
  &=&
  \prod_{i=1}^M \tau(\zeta_j/\xi_i)\tau(\zeta_j\xi_i)
  |\zeta_1,\ldots,\zeta_N ; \xi_1,\ldots,\xi_M \rangle^{(i)}.
\end{eqnarray}
Then the energy levels of the excited states are written as
\begin{equation}
  H_{\mbox{\scriptsize XXZ}}|1,\ldots,1;\xi_1,\ldots,\xi_M \rangle^{(i)}
  =
  \mu^{(i)}(r;\xi_1,\ldots,\xi_M)
  |1,\ldots,1;\xi_1,\ldots,\xi_M \rangle^{(i)},
\end{equation}
\begin{equation}
  \mu^{(i)}(r;\xi_1,\ldots,\xi_M) =
  (-1)^{1-i}\epsilon(r)
  + \sum_{i=1}^M
  \frac{2K}{\pi}\sinh\frac{\pi K'}{K}
  \mbox{dn}\left(\frac{2K}{\pi}\theta_i \right).
\end{equation}
We have parameterized $\xi=-ie^{i\theta}$.
Such a structure of the energy levels is 
similar to the quasi-particles structure pointed out
by Faddeev and Takhtajan in the XXX model \cite{FT}.
We conclude that our method gives the vacuum and the excited states.

\setcounter{equation}{0}
\section{Thermodynamic Limit}
\label{thermo}
We have constructed the eigenstates of the XXZ model
satisfying the turning symmetry in terms of the VOs.
With the aid of their free field realizations
one can compute their $2^N$-dimensional vector expressions.
However they contain the complex integrals over $N$ variables,
because a component of the VO $\Phi_+(\zeta)$
is represented in the integral form
(see the definition (\ref{F+})).
To avoid this difficulty, we notice that, in the bulk part,
the boundary q-KZ equation requests
only the commutation relation of the VOs (\ref{com_VO}).
This fact implies that
the vacuum state of the model can be obtained
without the representation theoretical meanings of the VOs.
In this section
we decompose the vacuum into a huge product of the R-matrices
and find its asymptotic form
applying the Baxter's formula on the corner transfer matrix
\cite{CBMS,B}.

\subsection{Half-Infinite Chain Limit}
\label{half}
To calculate magnetizations
we begin with the dual eigenstates of the transfer matrix.
We write them as
\begin{eqnarray}
  \label{z*}
  \hspace*{-1em}
  {}^{(i)}\langle \zeta_N,\ldots,\zeta_1|
  &=&
  {}^{(i)}\langle W^*|
  \Phi^*(\zeta_N)\otimes\cdots\otimes\Phi^*(\zeta_1)
  |V^*\rangle^{(i)}
  \nonumber \\
  &=&
  \sum_{\epsilon_1,\ldots,\epsilon_N}
  {}^{(i)}\langle W^*|
  \Phi_{\epsilon_N}^*(\zeta_N)\otimes\cdots\otimes
  \Phi_{\epsilon_1}^*(\zeta_1)
  |V^*\rangle^{(i)}
  (v_{\epsilon_N}^*\otimes\cdots\otimes v_{\epsilon_1}^*),
\end{eqnarray}
where the boundary states
${}^{(i)}\langle W^*|$ and $| V^*\rangle^{(i)}$
are defined by the relations,
\begin{eqnarray}
  \label{LW*}
  \sum_{\epsilon'} {}^{(i)}\langle W^*|
  \Phi_{\epsilon'}^*(\zeta^{-1}) K_{\epsilon'}^\epsilon(\zeta)
  &=&
  \Lambda_W^{(i)}(\zeta;r) {}^{(i)}\langle W^*| \Phi_\epsilon^*(\zeta),
  \\
  \label{LV*}
  \sum_{\epsilon'} \bar{K}_\epsilon^{\epsilon'}(\zeta)
  \Phi_{\epsilon'}^*(\zeta)|V^*\rangle^{(i)}
  &=&
  \Lambda_V^{(i)}(\zeta;r^{-1}) \Phi_\epsilon^*(\zeta^{-1})
  |V^*\rangle^{(i)}.
\end{eqnarray}
These dual eigenstates form an $N$-fold dual tensor space
$V_N^*\otimes\cdots\otimes V_1^*$
and satisfy
\begin{equation}
  {}^{(i)}\langle \zeta_1,\ldots,\zeta_N|
  T_j(\zeta_1,\ldots,\zeta_N) =
  \Lambda^{(i)}(\zeta_j;r){}^{(i)}\langle \zeta_1,\ldots,\zeta_N|.
\end{equation}
Let us explain why the dual states are given by the equation (\ref{z*}).
Consider the half-infinite chain limit.
One of the boundaries then vanishes and
the eigenstates are redefined by 
\begin{eqnarray}
  |\ldots,\zeta_N,\ldots,\zeta_1\rangle^{(i)}
  &=&
  \cdots\otimes
  \Phi(\zeta_N)\otimes\cdots\otimes\Phi(\zeta_1)
  |W\rangle^{(i)},
  \\
  {}^{(i)}\langle\zeta_1,\ldots,\zeta_N,\ldots|
  &=&
  {}^{(i)}\langle W^*|
  \Phi^*(\zeta_1)\otimes\cdots\otimes\Phi^*(\zeta_N)
  \otimes\cdots.
\end{eqnarray}
As an example
we introduce an operator $\mathcal{O}$ on site $m$
whose matrix elements are defined by
$\mathcal{O}.v_{\epsilon} =
\sum_{\epsilon'_m}\mathcal{O}_{\epsilon}^{\epsilon'} v_{\epsilon'}$
and calculate the vacuum expectation value of $\mathcal{O}$.
Use the unitarity relation of the VOs
$\sum_\epsilon \Phi_\epsilon^*(\zeta)\Phi_\epsilon(\zeta) = \mbox{id}$
\cite{CBMS}
and take care of the order of the $N$-fold tensor product
of two-dimensional vector spaces.
Then the vacuum expectation is written as
\begin{eqnarray}
  \langle\mathcal{O}\rangle^{(i)}
  &=&
  \sum_{\epsilon'_m,\epsilon_m,\ldots,\epsilon_1}
  {}^{(i)}\langle W^*|
  \Phi_{\epsilon_1}^*(\zeta_1)\cdots \Phi_{\epsilon_m}^*(\zeta_m)
  \mathcal{O}_{\epsilon_m}^{\epsilon'_m}
  \Phi_{\epsilon'_m}(\zeta_m)\cdots \Phi_{\epsilon_1}(\zeta_1)
  |W\rangle^{(i)}.
\end{eqnarray}
This result is the same as that in the half-infinite XXZ spin chain
except for the normalization \cite{JKKKM1}.

\subsection{Baxter's Formula}
We consider the thermodynamic limit of the vacuum.
We must deal with the infinite product of the VOs.
Note that, in the bulk part of the vacuum state,
only the commutation relation (\ref{com_VO})
has been used.
This relation is obtained from
the VO represented as the infinite product of the R-matrices.
Hence the VOs in our method
are replaced with a product of the R-matrices:
\begin{equation}
  \cdots\otimes\Phi(\zeta_B)\otimes\Phi(\zeta_B)\otimes\cdots
  =
  \cdots\otimes(\cdots\widehat{R}(\zeta_B)\widehat{R}(\zeta_B))\otimes
  (\cdots\widehat{R}(\zeta_B)\widehat{R}(\zeta_B))\otimes\cdots,
\end{equation}
where $\widehat{R}(\zeta)=R(\zeta)P$
and $\zeta_B$ stands for the spectral parameter in the balk part.
To treat such a huge group of the R-matrices
we employ the idea of corner transfer matrix,
which was originally introduced by Baxter
to calculate the correlation functions \cite{B}.
It is defined by
\begin{equation}
  A(\zeta_B) =
  \widehat{R}(\zeta_B)\otimes(\widehat{R}(\zeta_B)\widehat{R}(\zeta_B))
  \otimes\cdots\otimes
  (\widehat{R}(\zeta_B)\widehat{R}(\zeta_B)\cdots\widehat{R}(\zeta_B))
  \otimes\cdots.
\end{equation}
Baxter discovered that
the corner transfer matrix is
drastically simplified in the infinite chain limit;
\begin{equation}
  \label{A}
  A(\zeta_B) = \zeta_B^{-D}.
\end{equation}
Here $D$ is an operator independent of $\zeta_B$
and has the property
$\xi^{-D}\Phi(\zeta)\xi^D = \Phi(\zeta/\xi)$
(see Chapter 4 in \cite{CBMS}).
This asymptotic identify is derived from the Yang-Baxter equation (\ref{YBE})
and also understood 
in a framework of the crystal base \cite{K,KKMMNN}.
The identity (\ref{A}) is referred as the Baxter's formula.
We remark that
the ground states make no contribution to the Baxter's formula.
Here the ground states mean the completely antiferromagnetic states
where $+$ and $-$ alternatively appear at each edges of the R-matrices.

By virtue of the Baxter's formula
we can simplify the infinite product of the VOs.
The R-matrices of the VO at the infinitely far point are
in the ground states \cite{CBMS,B}.
In the same manner the R-matrices belonging to the VOs at 
sufficiently far sites are certainly in the ground states.
This reasoning implies that the upper triangle sector
of a huge group is in the ground states.
Notice that the R-matrices in the ground states are omitted.
Then the Baxter's formula gives
\begin{equation}
  \cdots\Phi(\zeta_B)\otimes\Phi(\zeta_B)\Phi_\epsilon(\zeta)
  |W\rangle^{(i)}
  =
  \zeta_B^{-D}\Phi_\epsilon(\zeta)|W\rangle^{(i)}.
\end{equation}
The VO at a boundary remains
because the defining relation (\ref{LW})
requests the explicit expressions of the VOs.

Next we consider the product of the VOs near another boundary.
Recall that the vacuum state satisfies the turning symmetry.
Such state is realized by the product of
the R-matrices and the $90^\circ$-rotated matrices.
Therefore the turning symmetry tells that
the product near ${}^{(i)}\langle V|$
is related to that near $|W\rangle^{(i)}$
through the crossing symmetry (\ref{cross}).
Thus, the vacuum state in the infinite chain limit
is simplified as follows,
\begin{eqnarray}
  |\zeta,\zeta_B \rangle^{(i)}
  &=&
  {}^{(i)}\langle V|
  \cdots\otimes\Phi(-q^{-1}\zeta_B)\otimes\Phi(-q^{-1}\zeta_B)
  \Phi_{-\epsilon}(-q^{-1}\zeta)
  \nonumber \\
  && \qquad\times
  \cdots\otimes\Phi(\zeta_B)\otimes\Phi(\zeta_B)
  \Phi_\epsilon(\zeta)
  |W \rangle^{(i)}
  \nonumber \\
  &=&
  {}^{(i)}\langle V|
  (-q^{-1}\zeta_B^{-1})^{-D}
  \Phi_{-\epsilon}(-q^{-1}\zeta)
  \zeta_B^{-D}
  \Phi_\epsilon(\zeta)
  |W \rangle^{(i)}
  \nonumber \\
  &=&
  {}^{(i)}\langle V|
  \Phi_{-\epsilon}(\zeta\zeta_B)(-q)^D \Phi_\epsilon(\zeta)
  |W \rangle^{(i)}.
\end{eqnarray}
By the definition
the spectral parameter $\zeta_B$
of the rotated VOs remains unchanged \cite{CBMS}.

Similarly we get the dual state of the vacuum
in the infinite chain limit,
\begin{equation}
  {}^{(i)}\langle \zeta,\zeta_B| =
  {}^{(i)}\langle W^*|
  \Phi_\epsilon^*(\zeta\zeta_B)(-q)^D \Phi_{-\epsilon}^*(\zeta)
  |V^* \rangle^{(i)}.
\end{equation}
This infinite chain limit is regarded as the thermodynamic limit
because boundaries on both sides survive.
We thus have simplified the expressions of the vacuum states
in the thermodynamic limit.

\setcounter{equation}{0}
\section{Spontaneous Magnetization in the Finite XXZ \\ Spin Chain}
\label{spon}
\subsection{Free Field Representations of the Vertex Operators}
To calculate the spontaneous magnetization at a boundary,
we prepare the free field representations of the VOs
and realize the boundary states.

The highest weight representations of $U_q(\widehat{sl}_2)$ with level $1$
are realized as the Fock spaces,
which consist of
the bosonic operators $a_k$ ($k\in \mathbb{Z}/\{0\}$) 
and the zero-modes $\Lambda_i$, $\alpha$ ($i=0,1$),
\begin{eqnarray}
  V(\Lambda_i) &=&
  \mathbb{C}[a_{-1},a_{-2},\ldots]|0\rangle \otimes
  (\oplus_{n\in\mathbb{Z}} e^{\Lambda_i+n\alpha}),
  \\
  V^*(\Lambda_i) &=&
  \langle 0|\mathbb{C}[a_1,a_2\ldots] \otimes
  (\oplus_{n\in\mathbb{Z}} e^{-\Lambda_i-n\alpha}).
\end{eqnarray}
Here the bosonic operators satisfy the following commutation relation,
\begin{eqnarray}
  [a_k,a_l] = \frac{[2k][k]}{k}\delta_{k+l,0},
  \qquad
  [n] = \frac{q^n -q^{-n}}{q-q^{-1}}.
\end{eqnarray}
The action of the zero-mode is defined by
\begin{equation}
  e^\gamma .e^\beta = e^{\beta+\gamma},
  \qquad
  z^\partial .e^\beta = z^{[\partial,\beta]} e^\beta,
\end{equation}
for $V(\Lambda_i)$ and 
\begin{equation}
  \label{V*}
  e^\beta .e^\gamma = e^{\beta+\gamma},
  \qquad
  e^\beta .z^\partial = e^\beta z^{[\beta,\partial]},
\end{equation}
for $V^*(\Lambda_i)$.
The zero-mode $\Lambda_i$ is related with $\Lambda_1=\Lambda_0+\alpha/2$.
The operator $\partial$ is determined
by the commutation relations
$[\partial,\alpha] = 2$ and $[\partial,\Lambda_0] = 0$.
The zero-modes $\Lambda_i$ and $\alpha$ are regarded as
the fundamental weights and one of the simple roots of $\widehat{sl}_2$,
respectively.
We write
\begin{equation}
  |0 \rangle^{(i)} = |0\rangle \otimes e^{\Lambda_i},
  \qquad
  {}^{(i)}\langle 0| = \langle 0| \otimes e^{-\Lambda_i}.
\end{equation}

The VOs of type I are bosonized as follows,
\begin{eqnarray}
  \Phi_-^{(1-i,i)}(\zeta)
  &=&
  e^{P(\zeta^2)}e^{Q(\zeta^2)}\otimes
  e^{\alpha/2}(-q^3\zeta^2)^{\frac{\partial+i}{2}}\zeta^{-i},
  \\
  \label{F+}
  \Phi_+^{(1-i,i)}(\zeta)
  &=&
  \oint_C \frac{dw}{2\pi\sqrt{-1}}
  \frac{(1-q^2)w\zeta}{q(w-q^2\zeta^2)(w-q^4\zeta^2)}
  :\Phi_-^{(1-i,i)}(\zeta)X(w):,
  \\
  X(w) &=&
  e^{R(w)}e^{S(w)}\otimes e^{-\alpha}w^{-\partial}.
\end{eqnarray}
The integration contour $C$ encircles around $0$ in such a way that
$|q^4\zeta^2|<|w|<|q^2\zeta^2|$.
The normal-ordering product $:$ $:$ plays the same role as
that in the conventional bosonic operators.
The ``primary fields'' are
\begin{eqnarray}
  &&
  P(z) = \sum_{k=1}^\infty \frac{a_{-k}}{[2k]} q^{7k/2}z^k,
  \qquad
  Q(z) = -\sum_{k=1}^\infty \frac{a_k}{[2k]} q^{-5k/2}z^{-k},
  \\ &&
  R(w) = -\sum_{k=1}^\infty \frac{a_{-k}}{[k]} q^{k/2}z^k,
  \qquad
  S(w) = \sum_{k=1}^\infty \frac{a_k}{[k]} q^{k/2}z^{-k}.
\end{eqnarray}

\subsection{Boundary States}
\label{bound}
We find the boundary states
using the free field representations of the VOs.
First we realize two boundary states
$|W\rangle^{(i)}$ and ${}^{(i)}\langle W^*|$.
The commutation relations
for $[A,[A,B]]\in\mathbb{C}$,
\begin{equation}
  e^A e^B = e^{\frac{1}{2}[A,[A,B]]}e^{[A,B]}e^B e^A,
  \qquad
  e^B e^A = e^A e^B e^{[B,A]}e^{\frac{1}{2}[[B,A],A]},
\end{equation}
and the formula
\begin{equation}
  \exp\left(
    \sum_{k=1}^\infty \frac{[mk]}{[nk]}z^k
  \right) =
  \frac{(q^{n+m}z;q^{2n})_\infty}{(q^{n-m}z;q^{2n})_\infty},
\end{equation}
are useful.
From the relation (\ref{LW}) with $\epsilon=-$
and the relation (\ref{LW*}) with $\epsilon=+$,
we derive the boundary states
$|W\rangle^{(i)}$ and ${}^{(i)}\langle W^*|$ as
\begin{eqnarray}
  \label{rW}
  |W\rangle^{(i)}
  &=&
  \exp\left(
    -\frac{1}{2}\sum_{k=1}^\infty
    \frac{kq^{6k}}{[2k][k]} a_{-k}^2
    + \sum_{k=1}^\infty f_k^{(i)} a_{-k} \right)
  |0\rangle^{(i)},
  \\
  \label{rW*}
  {}^{(i)}\langle W^*|
  &=&
  {}^{(i)}\langle 0|
  \exp\left(
    -\frac{1}{2}\sum_{k=1}^\infty
    \frac{kq^{-2k}}{[2k][k]} a_k^2
    + \sum_{k=1}^\infty g_k^{(i)} a_k \right).
\end{eqnarray}
Here the functions $f_k^{(i)}$ and $g_k^{(i)}$ are given by
\begin{eqnarray}
  f_k^{(i)}
  &=&
  \frac{q^{5k/2}}{[2k]}
  \left(-\theta_k(1-q^k)+(-1)^{1-i}q^{(1-2i)k}r^{(1-2i)k}\right),
  \\
  g_k^{(i)}
  &=&
  \frac{q^{-3k/2}}{[2k]}
  \left(\theta_k(1-q^k)+(-1)^{1-i}q^{-(1-2i)k}r^{(1-2i)k}\right),
\end{eqnarray}
with $\theta_k=(1+(-1)^k)/2$.
For the relations with other components we must verify
these realizations.
As an example we consider
the relation (\ref{LW}) with $\epsilon=+$ in the case $i=0$.
The realization (\ref{rW}) reduces 
the relation (\ref{LW}) to the following integral equation,
\begin{equation}
  \oint_{C} \frac{w^{-1}dw}{2\pi\sqrt{-1}}
  \frac{(1-r\zeta^2)(w^2-q^6)(w-rq^4)}
  {(w-q^2\zeta^2)(w-q^4\zeta^2)(w-q^4\zeta^{-1})}
  e^{R(w)+R(q^6 w^{-1})} |0\rangle^{(0)}
  =
  (\zeta \leftrightarrow \zeta^{-1}) |0\rangle^{(0)}.
\end{equation}
In the left hand side
we change integration variable as $w\rightarrow q^6 w^{-1}$ and
take the average of the original and the transformed terms.
We then get the symmetrical integral expression
for the spectral parameter $\zeta$ and $\zeta^{-1}$.

Other boundary states are found
from the following identities,
\begin{equation}
  \label{rel_B}
  {}^{(i)}\langle V| = {}^{(i)}\langle W^*|(-q)^{-D},
  \qquad
  |V^*\rangle^{(i)} = (-q)^{-D} |W\rangle^{(i)}.
\end{equation}
They are proved by making use of the equation
$K_\epsilon^{\epsilon'}(\zeta;r) =
\Lambda(\zeta;r) K_{-\epsilon}^{-\epsilon'}(\zeta;r^{-1})$.
For instance,
putting the first identity of (\ref{rel_B})
into the relation (\ref{LW*}),
we have
\begin{eqnarray}
  \lefteqn{\sum_{\epsilon'} {}^{(i)}\langle W^*|
    \Phi_{\epsilon'}^*(\zeta^{-1}) K_{\epsilon'}^\epsilon(\zeta)}
  \nonumber \\
  &=&
  \Lambda(\zeta;r)
  \sum_{\epsilon'} {}^{(i)}\langle V|(-q)^D
  \Phi_{-\epsilon'}(-q^{-1}\zeta^{-1})
  K_{-\epsilon'}^{-\epsilon}(\zeta;r^{-1})
  \nonumber \\
  &=&
  \Lambda(\zeta;r)\Lambda_V^{(i)}(\zeta;r^{-1})
  {}^{(i)}\langle V| \Phi_{-\epsilon}(\zeta) (-q)^D
  \nonumber \\
  &=&
  \Lambda_W^{(i)}(\zeta;r)
  {}^{(i)}\langle W^*| \Phi_\epsilon^*(\zeta).
\end{eqnarray}
The defining relations (\ref{LW}) and (\ref{LW*}) are
reproduced by the relations (\ref{LV*}) and (\ref{LV})
through the identities (\ref{rel_B}) respectively.

We explain briefly why factors $\Lambda_W^{(i)}(\zeta;r)$ and
$\Lambda_V^{(i)}(\zeta;r)$ appear
in the defining relation of the boundary states.
In section \ref{half} we have formulated the dual eigenstates
by comparing the correlation function in our method with
the already known result.
This argument is valid only if
the states $|W\rangle^{(i)}$ and ${}^{(i)}\langle W^*|$ are
the completely same as those in the half-infinite XXZ spin chain.
Thus, factors $\Lambda^{(i)}(\zeta;r)$ are attached to these states
(see \cite{JKKKM1}).
The factors in all defining relations
are determined through the relations (\ref{rel_B}).

\subsection{Spontaneous Magnetization}
We are in a position to calculate the spontaneous magnetization
in the finite XXZ spin chain.
By the Baxter's formula (\ref{A}) and the identities (\ref{rel_B}),
the vacuum states are written as
\begin{eqnarray}
  |\zeta_\pm\rangle^{(i)}
  &=&
  {}^{(i)}\langle V|\Phi_{-\epsilon}(\zeta_-)(-q)^D
  \Phi_\epsilon(\zeta_+)|W\rangle^{(i)}
  \nonumber \\
  &=&
  {}^{(i)}\langle W^*|\Phi_{-\epsilon}(-q^{-1}\zeta_-)
  \Phi_\epsilon(\zeta_+)|W\rangle^{(i)},
  \\
  {}^{(i)}\langle \zeta_\pm|
  &=&
  {}^{(i)}\langle W^*|\Phi_\epsilon^*(\zeta_-)(-q)^D
  \Phi_{-\epsilon}^*(\zeta_+)|V^*\rangle^{(i)}
  \nonumber \\
  &=&
  {}^{(i)}\langle W^*|\Phi_{-\epsilon}(-q^{-1}\zeta_-)
  \Phi_\epsilon(\zeta_+)|W \rangle^{(i)}.
\end{eqnarray}
Therefore only the following quantity is of our main concern,
\begin{equation}
  P_\epsilon^{(i)}(\zeta_\pm^2;r)
  =
  {}^{(i)}\langle W^*|\Phi_{-\epsilon}(-q^{-1}\zeta_-)
  \Phi_\epsilon(\zeta_+)|W \rangle^{(i)}.
\end{equation}
Noting the order of the sites in the dual state
we have the magnetization at a boundary,
\begin{equation}
  \langle \sigma^z \rangle^{(i)}(r) =
  \frac{(P_+^{(i)}(1;r))^2 - (P_-^{(i)}(1;r))^2}
  {(P_+^{(i)}(1;r))^2 + (P_-^{(i)}(1;r))^2}.
\end{equation}
By the free field representations of the VOs
we get the following bosonic expression of $P_\epsilon^{(i)}(z_\pm;r)$,
\begin{equation}
  P_\epsilon^{(i)}(z_\pm;r)
  =
  \epsilon \frac{(q^4;q^4)_\infty}{(q^6;q^4)_\infty}
  \oint_{C_\epsilon^{(i)}} \frac{w^{1-i}dw}{2\pi\sqrt{-1}}
  \frac{q^{2i}(1-q^2)
    z_-^{(1+i)/2}z_\epsilon^{i/2}z_+^{i/2}
    Y^{(i)}(z_\pm,w;r)}
  {(w-z_-)(w-q^2 z_\epsilon)(w-q^4 z_+)}.
\end{equation}
The integration contours $C_+^{(i)}$ and $C_-^{(i)}$ encircle clockwise
around $0$
in such a way that $|q^4 z_\pm|<|w|<|q^2 z_\pm|$
and $|q^2 z_\pm|<|w|<|z_\pm|$,
respectively.
$Y^{(i)}(z_\pm,w;r)$ is the inner product of normal-ordered bosonic operators
given by
\begin{equation}
  Y^{(i)}(z_\pm,w;r) =
  {}^{(i)}\langle W^*|
  \exp\left(\sum_{k=1}^\infty a_{-k} G_k \right)
  \exp\left(-\sum_{k=1}^\infty a_k F_k \right)
  |W \rangle^{(i)},
\end{equation}
where the functions $F_k$ and $G_k$ are
\begin{eqnarray}
  F_k &=&
  \frac{q^{-5k/2}}{[2k]}(z_+^{-k}+q^{2k}z_-^{-k})
  - \frac{q^{k/2}}{[k]}w^{-k},
  \\
  G_k &=&
  \frac{q^{7k/2}}{[2k]}(z_+^k+q^{-2k}z_-^k)
  - \frac{q^{k/2}}{[k]}w^k.
\end{eqnarray}
Since both boundary states (\ref{rW}) and (\ref{rW*})
contain the quadratic terms
of bosonic operators on the exponential function,
the commutation relations among bosons are not useful.
First we insert twice the completeness relation,
\begin{equation}
  \mbox{id}_{V(\Lambda_i)} =
  \int \mathcal{D}|\xi|^2
  \exp\left(-\sum_{k=1}^\infty\frac{k}{[2k][k]}|\xi_k|^2\right)
  |\xi\rangle^{(i)}{}^{(i)}\langle \bar{\xi}|,
\end{equation}
between the boundary states and the exponential operators.
Here the integration is taken over the entire complex plane
and the measure $\mathcal{D}|\xi|^2$ means
the infinite product $\prod_{k=1}^\infty d|\xi_k|^2$
except for the normalization.
The coherent states $|\xi\rangle^{(i)}$ and ${}^{(i)}\langle\bar{\xi}|$
are given by
\begin{eqnarray}
  |\xi\rangle^{(i)}
  &=&
  \exp\left(\sum_{k=1}^\infty
    \frac{k}{[2k][k]}\xi_k a_{-k}
  \right)|0\rangle^{(i)},
  \\
  {}^{(i)}\langle \bar{\xi}|
  &=&
  {}^{(i)}\langle 0|
  \exp\left(\sum_{k=1}^\infty
    \frac{k}{[2k][k]}\bar{\xi}_k a_k
  \right).
\end{eqnarray}
They clearly satisfy the definition of the coherent states;
\begin{equation}
a_k|\xi\rangle^{(i)} = \xi_k|\xi\rangle^{(i)},
{}^{(i)}\langle \bar{\xi}|a_{-k} =
{}^{(i)}\langle \bar{\xi}|\bar{\xi}_k.  
\end{equation}

Next we use the formula for the Gauss integration,
\begin{eqnarray}
  \lefteqn{\int\mathcal{D}|\vec{\xi}|^2
    \exp\left(
      -\frac{1}{2}\sum_{k=1}^\infty \frac{k}{[2k][k]}
      {}^t\vec{\xi}_k A_k \vec{\xi}_k
      + \sum_{k=1}^\infty {}^t\vec{\xi}_k B_k
    \right)}
  \nonumber \\
  &=&
  \exp\left(\frac{1}{2}\sum_{k=1}^\infty
    \frac{[2k][k]}{k}\:
    {}^t\! B_k A_k^{-1} B_k
  \right).
\end{eqnarray}
For $Y^{(i)}(z_\pm,w;r)$
$A_k$ is an invertible $4\times 4$ matrix
and $B_k$ is a four-component vector.
Then the inner product $Y^{(i)}(z_\pm,w;r)$ is reduced to
\begin{eqnarray}
  Y^{(i)}(z_\pm,w;r)
  &=&
  \exp\Biggl(-\sum_{k=1}^\infty \frac{[2k][k]}{k}
  \frac{1}{1-q^{4k}}
  \biggl\{
    \frac{1}{2}q^{-2k}F_k^2 + q^{4k}F_k G_k
    +\frac{1}{2}q^{6k}G_k^2
  \nonumber \\
  && \qquad
  -(g_k^{(i)}-q^{-2k}f_k^{(i)})F_k
  +(f_k^{(i)}-q^{6k}g_k^{(i)})G_k
  \nonumber \\
  && \qquad
  +\frac{1}{2}q^{-2k}f_k^{(i)2}
  -f_k^{(i)}g_k^{(i)}
  +\frac{1}{2}q^{6k}g_k^{(i)2}
  \biggr\} \Biggr).
\end{eqnarray}
Taking out terms depending on the parameter $\epsilon$, $i$ and $w$,
we have the following integral expression,
\begin{equation}
  \label{Pi}
  P_\epsilon^{(i)}(z;r)
  =
  \oint_{C_\epsilon^{(i)}}
  \frac{\epsilon dw}{2\pi\sqrt{-1}}
  \frac{1-wzq^{-2}}{1-wrq^{-2}}
  \frac{\Theta_{q^4}(q^2 w^2)}
  {\Theta_{q^2}(wz)\Theta_{q^2}(wz^{-1})}
\end{equation}
Here we have put $z=z_+=z_-$,
since we have no poles that degenerate with this identification.
The integration contour $C_+^{(0)}$ picks out the points
$q^{2n}z, q^{2n}z^{-1}$ for $n=2,3,\ldots$
and the contour $C_-^{(0)}$ the same points for $n=1,2,\ldots$.
The contour $C_\pm^{(1)}$ encircles around
not only the points in $C_\pm^{(0)}$
but the point $q^2 r^{-1}$.
With the aid of the quasi-periodicity
property $\Theta_p(pz)=-z^{-1}\Theta_p(z)$
the integral expression (\ref{Pi}) is easily calculated.
We thus arrive at the following results:
\begin{eqnarray}
  &&
  P_+^{(0)}(1;r) =
  \sum_{l=1}^\infty
  \frac{(-q^2)^l (1-r)^2}{(1-q^{2l}r)^2},
  \\
  &&
  P_-^{(0)}(1;r) = -(1+P_+^{(0)}(1;r)),
  \qquad
  P_\epsilon^{(1)}(1;r) = -P_\epsilon^{(0)}(1;r^{-1}).
\end{eqnarray}
Here
we have normalized the result such that
the sum of $P_+$ and $P_-$ equals $-1$.
With $P^{(i)}(1;r)$ the magnetization at a boundary is given by
\begin{equation}
  \langle\sigma^z\rangle^{(i)}(r) =
  -\frac{1+2P^{(i)}(1;r)}{1+2P^{(i)}(1;r)+2(P^{(i)}(1;r))^2}.
\end{equation}

We see that the magnetization survives
even if the boundary magnetic fields vanish.
Substituting $-1$ into the parameter $r$ (see (\ref{h})),
we obtain the spontaneous magnetization in the finite XXZ spin chain
as follows,
\begin{equation}
  \langle\sigma^z\rangle =
  \langle\sigma^z\rangle^{(0)}(-1) =
  \langle\sigma^z\rangle^{(1)}(-1) =
  -\frac{2(q^4;q^4)_\infty^4}
  {(q^2;q^2)_\infty^8 + (-q^2;q^2)_\infty^8}.
\end{equation}
The spontaneous magnetization in the half-infinite XXZ spin chain
is known to be $-(q^2;q^2)_\infty^4/(-q^2;q^2)_\infty^4$
\cite{JKKKM1}.
Denoting it by $\langle\sigma^z\rangle_\infty$
we have
\begin{equation}
  \label{result}
  \langle\sigma^z\rangle =
  2\langle\sigma^z\rangle_\infty
  -2\langle\sigma^z\rangle_\infty^3
  +2\langle\sigma^z\rangle_\infty^5
  -\cdots.
\end{equation}
The relation (\ref{result}) shows that
the spontaneous magnetization in the finite chain
is roughly twice as large as that in the half-infinite chain
when $q$ is close to $-1$
or when the anisotropy is small.
We conclude that, in the XXZ model,
the anisotropy generates the spontaneous magnetization
and the finiteness of the model enhances it.

\setcounter{equation}{0}
\section{Summary and Discussion}
We have studied the finite XXZ spin chain with boundaries.
The vacuum state is derived from the boundary q-KZ equation
and is composed by the type I VOs.
Apart from the representation theory
we have interpreted a large product of the VOs
as a huge group of the R-matrices
and have simplified the vacuum state in the thermodynamic limit
using the Baxter's formula.
Simple expressions of the states enable us to calculate
the spontaneous magnetization in the finite XXZ spin chain.
The spontaneous magnetizations in the finite chain
and in the half-infinite chain
are related by (\ref{result}).
It has been made clear that
the vertex operator approach is valid for the analysis of
the finite XXZ spin chain.

We have constructed the excited states
by including the type II VOs in the vacuum
and have shown that
the energy levels are similar to the quasi-particle structure
in the XXX model \cite{FT}. 
Such structure is known to appear
in the infinite chain limit.
We bypass arguments whether there exist these energy levels
in a finite chain,
since we attach importance to the thermodynamic limit.

In the paper we have considered only the diagonal K-matrix.
It is interesting to solve the same problem
when the K-matrix has non-diagonal elements.
The non-diagonal K-matrix generates
not only the boundary magnet fields
but the boundary flows of spinons \cite{VG}.
Since the sum of spins is not preserved
we might expect a novel phenomenon.
To treat the non-diagonal K-matrix
one needs the complex integrals,
because each defining relation of the boundary states
is expressed by an equation such that two VOs are combined into one.
Such equations can not be realized in terms of only bosonic operators.

We have regarded the infinite product of the VOs
as the thermodynamic limit
since both boundaries remain.
The thermodynamic limit should be taken
by fixing a ratio of two infinite parameters.
In our method we may choose
the ratio between the number of the type II VOs
and the number of the sites.
In the paper, however,
we have not given attention to the fixed parameter,
because only the vacuum state is considered.
In return this choice implies that
one must treat the infinite number of the type II VOs
when one compute the thermodynamic limit of the excited states.
It is difficult
to find the behavior of the infinite product of the type II VOs
by their free field representations.
A way to solve the difficulty is to find
the recursion relations among the excited states.

Korepin proposed a method to calculate
the correlation functions in the integrable systems \cite{QISM}.
There, the recursion relations are
derived from the algebraic Bethe ansatz.
There may exist a similarity between the recursion relations by our method
and those by the algebraic Bethe ansatz.
Referring to this similarity
one can realize the Korepin's method by the vertex operator approach.
This expectation is supported by a fact
that some combinations of the type I VOs and the type II VOs
correspond to the L-operator in the Yang Baxter equation \cite{Mi}.

\section*{Acknowledgment}
The authors thank to H. Ujino, M. Shiroishi, Y. Kajinaga and T. Tsuchida
for useful discussions.

\end{document}